  \def\etc {{\it etc.}}

  \def\wt    {$\omega(\theta)\ $}

  \def\Aw8   {$A_{\omega}^{\delta=0.8}$\ }

\documentstyle[11pt,aas2pp4]{article}

\slugcomment{Submitted to ApJ.}

\lefthead{Infante et al.}
\righthead{Faint Interacting Galaxies}

\begin{document}

\title{Strong Clustering of Faint Galaxies at Small Angular Scales}

\author{Leopoldo Infante}
\affil{Departamento de Astronom\'{\i}a y Astrof\'{\i}sica,
P. Universidad Cat\'olica de Chile, Casilla 104, Santiago 22, Chile}

\author{Du\'\i lia F. de Mello\altaffilmark{1}} 
\affil{Observat\'orio Nacional - DAN, RJ, 20921-400 Brazil}

\and

\author{Felipe Menanteau}
\affil{Departamento de Astronom\'{\i}a y Astrof\'{\i}sica,
P. Universidad Cat\'olica de Chile, Casilla 104, Santiago 22, Chile}

\altaffiltext{1}{CNPq Fellow}

\begin{abstract}

The 2-point angular correlation function of galaxies, \wt, has been computed on
equatorial fields observed with the CTIO 4m prime focus, within a total area of
2.31 deg$^2$.  In the magnitude range $19\le m_R \le 21.5$, corresponding to
$<z>\approx 0.35$, we find an excess of power in \wt at scales  $2''\le\theta
\le6''$ over what would be expected from an extrapolation of \wt
measured at larger
$\theta$. The significance of this excess is $\approx 5\sigma$. At larger
scales, $6''< \theta \le 24''$,  the amplitude of \wt is 1.6 times smaller than
the standard no evolutionary model. At these scales there is remarkable
agreement between the present data and Infante \& Pritchet (1995). 

At large angular scales ($6''< \theta \le 24''$) the data is best described by a
model where clustering evolution in $\xi(r,z)$ has taken place. Strong
luminosity evolution cannot be ruled out with the present data. At smaller
scales, $2''\le \theta \le 6''$, our data are formally fit by models where
$\epsilon=-2.4 (\Omega=0.2, r_o=5.1h^{-1}$Mpc) or $r_o = 7.3h^{-1}$Mpc
$(\Omega=0.2, \epsilon=0)$. If the mean redshift of our sample is 0.35 then our
data show a clear detection of the scale ($\approx 19h^{-1}kpc$) where the
clustering evolution approaches a highly non linear regime, i.e., $\epsilon \le
0$.

The rate at which galaxies merge has been computed. If this rate is proportional
to $(1+z)^m$, then $m=2.2 \pm 0.5$.

\end{abstract}
 

\section{Introduction}

The two point angular correlation function has been extensively used to study
the clustering properties of galaxies. At large angular separations, $\theta >
10''$, it is well established that faint galaxies are  less clustered  compared
to galaxies observed locally. Infante \& Pritchet (1995) using observed redshift
distributions and straightforward calculations with little model dependency
found  that predicted angular correlation amplitudes are about a factor of two
higher than those observed over a wide range from $20<m_B<24$. The blue galaxy
excess in the counts N(m), and the correlation deficiency, can be neatly
resolved by introducing a ``new'', weakly clustered population at low redshifts
of $0.2 < z < 1$ (Efstathiou et al. 1991; Infante \& Pritchet 1995). However,
luminosity-dependent evolution must be invoked in order to brighten up invisible
low-L galaxies at $z=0$ to $\sim L^*$ at $z \approx 0.3$, while still
maintaining the no-evolution shape of N(z) (Colless et al. 1990; Lilly et al.
1991; Broadhurst et al. 1988). 

At small angular scales, $\theta \le 10''$, the situation is different and there
is no consensus yet. A limited number of studies have been successful in
measuring the angular correlation function at small separations
($\theta<6\arcsec$, physical separation less than $\sim 20 h^{-1} kpc$ at
$z\approx0.3$)(Carlberg et al. 1994;   Neuschaefer et al. 1995; Woods et al.
1995). The main difficulty in such determinations is poor statistics on scales
$\le$ 10$''$.  The results range from a strong  excess   to a deficiency of
pairs with respect to an inward extrapolation of $\omega(\theta)\propto
\theta^{-0.8}$ at larger separations. This excess in \wt   may be due to a population of
``companions'' not present at the current epoch, or luminosity enhancement of
intrinsically faint galaxies in pairs.

In this letter we present the first clear detection of a break from a power law
observed at larger separations in the angular two point correlation function,
$\omega(\theta)$, at $\theta<6''$. We have measured \wt at separations 
$2''<\theta \le 68''$
on a 2.31 deg$^2$ field at $19\le m_R \le 21.5$ ($<z>\approx0.35$) from a recently
selected  catalog of faint galaxy  pairs and groups (de Mello et
al. 1996a and b, hereafter Paper I and III). Special care was taken in this work
to resolve galaxies at small separations.

\section{The Data}
The data set is comprised of 96 equatorial images of $15' \times 15'$ (0.44$''$
pixel$^{-1}$) making a total area of 5.25~deg$^{2}$ taken at the CTIO 4m prime
focus camera by the High-Z Supernovae Search Group (Leibundgut et al. 1995) over
two nights in March and November 1995.   The fields are all equatorial within $0^{h}
\le$ R.A. $\le 4^{h}$ and $10^{h} \le$ R.A. $\le 14^{h}$. Center field coordinates will be
given in a forthcoming paper. Five minute exposures through a redshifted B
filter which is almost equivalent to a regular Kron-Cousin  R filter (B/(z=0.4),
hereafter $m_{R}$) were sufficient to provide good quality images. After a
careful inspection of the images we decided to run our analysis on the best
images. Our selection criteria were based on seeing, low number of bright stars
and photometric conditions. The total effective angular area on which we have
computed the angular correlation function is 2.31~deg$^{2}$.

The observations, reductions, calibrations and selection criteria are 
described in Paper I. Coordinates of each pair/group is also given in
Paper I and III. A number of tests were conducted in order to determine 
detection performance and limiting magnitudes. We claim in Paper I 
that our catalog is 99\% complete at $m_R~<~22$, and that the two 
components of a pair separated by more than $2''$ were always detected
for $1.0''\leq seeing \leq 1.6''$.
Star/galaxy separation was done using Kron photometry (Kron 1980) and
the properties of the inverse first and second moments of the images
which gives a measure of intrinsic size and central compactness
(see Paper III for more details). Results were tested by eye inspection.

\section{Analysis and Results}

The angular correlation function, $\omega(\theta)$, was computed
using a method that is
more fully described in Infante (1994; see also Infante and Pritchet 1995,
hereafter IP95).  An artificial catalog of randomly distributed objects was
created, and areas contaminated by defects, bright stars, \etc\ were masked from
both the artificial catalog and the real catalog. The estimator used was

\begin{equation}
\omega (\theta )={{N_{gg}~N_r}\over{B~N_{gr}~(N_g-1)}}~-
{{N_{rr}~N_{r_1}}\over{N_{rr_{1}}~(N_r-1)}}
\end{equation}

\noindent
where $N_{gg}$ is the number of galaxy pairs in a given range of separations
summed over all CCD fields,
$N_{gr}$ is the number of random pairs using galaxies as centers, $N_g$ is the
number of galaxies, $N_r$ is the number of random objects, $N_{rr}$ is the
number of random pairs, and $N_{rr_{1}}$ is the same as $N_{rr}$ but an
independent set of random counts.  $B$ is the ``integral constraint'' correction
factor (e.g. Peebles 1980; Koo and Szalay 1984). At the relevant scales in this
paper, $\theta < 24''$, the uncertainties are limited by Poisson noise
rather than by variations in $B$.  We have used $B=1$ for our clustering at
small separation analysis.

The angular correlation function \wt was derived in a 2.31deg$^2$ area. Fig.
\ref{fig:w21.5} gives the angular correlation function at separations $\theta <
2'$, for the magnitude interval 19$\le m_{R}\le $21.5. Also, the canonical
$\omega$=2.3$\theta$$^{-0.8}$ , as derived in IP95, is given for comparison in
Fig. \ref{fig:w21.5} (thick solid line). 
The uncertainties, $\delta_{\omega}$, are 65\%  confidence intervals computed
using the bootstrap resampling method as described in Efron \& Tibshirani
(1986). Five hundred resamplings were carried out in each computation. We note
that Poisson errors are $\sim 2.5$ times smaller than bootstrap errors.  In most
data points the $2\sigma$ Poisson  error bars are smaller than the symbol.

Using least squares to fit \wt we obtain
$\omega=(2.06\pm0.02)\theta^{-0.79\pm0.02}$, in the range
$6''<~\theta~\le67.6''$. A K-S significance  test shows
that in this range, both  the canonical IP95 and \wt for our data, are drawn
from the same distribution with a 90\% significance level.

Table \ref{tab:excess} shows the number, N, of pairs at separations
$2''<\theta\le6''$ and  $6''<\theta\le24''$ for the real
and random catalogs. The fractional excess, 

\begin{equation}
\left({{\Delta N }\over{N}}\right)_{pairs} = \frac{\Delta \omega}{1+\omega},
\end{equation}

\noindent
in the  number of pairs with respect to the random catalog are  0.637 and 0.200,
respectively, while $\left({{\Delta N }\over{N}}\right)_{pairs}$ over what would
be expected from IP95 are   0.361 and -0.011, respectively. We also show in
Table \ref{tab:excess} the number of isolated groups, with  2, 3, 4, 5 and 6
members, found in our catalog and the number of groups found in the random
catalog.  

In Fig. \ref{fig:dn21.5} the fractional excess number of pairs,  $\left({{\Delta
N }\over{N}}\right)_{pairs}$, for $19\le m_R \le 21.5$ are plotted with respect
to the random catalog expectation (thin line), to IP95 data (thick line)
and to various models (dashed lines) as explained in the next section. At scales
$\theta \le 6''$ our data departs from the canonical \wt from IP95 by
$\approx 5 \sigma$. This excess power has not been detected in most previous
works, only Carlberg et al. (1994) report an excess in the number of pairs at
separations $< 6''$  and $V < 22.5 $.



\section{Discussion}

Are there any effects in the data or data reduction techniques that might in
principle cause an excess in \wt at small separations? First, we consider the
possibility that rich clusters of galaxies at $z\approx0.3$ in our fields might
significantly  enhance \wt at $2''<\theta\le6''$. From Table \ref{tab:excess}
the contribution of groups  with more than 5 members is 52 pairs corresponding
to a negligible $\left({{\Delta N }\over{N}}\right)_{pairs}=0.032$. Next, we
consider multiple detections of several faint spurious sources for each
physically distinct galaxy by our finding algorithm. For this purpose, all
images were inspected by eye, eliminating all possible sources of contamination,
namely, bright galaxies, scratches, bad pixels and cosmetic defects on the chip.
There is also the possibility of contaminating light. Galaxies that normally
would have been below the magnitude limit may be raised above it by the light of
the second object. Our photometry accounts for this effect.  We do an excision
of disturbing objects for crowded field photometry. Our algorithm takes the
median in rings and replaces deviant values with the median. And finally, we checked
for systematic effects on magnitude determination. Extensive simulations were
performed in order to test the detection and photometry of faint images as a
function of magnitude. No systematic effects were detected. See Paper I for
details regarding the simulations.

\subsection{Comparison with other Observations}

Several studies have investigated \wt of faint galaxies in a number of passbands
(IP95 and references therein; Neuschaefer et al. 1995; Shepherd et al. 1996). The
principal result from these surveys is that \wt follows a power law,
$\omega~=~A~\theta^{-\delta}$, where $0.7<\delta<0.9$ ($\delta=0.8$ is the
standard value), and A is the amplitude scaling with magnitude. In this work we
have taken IP95 as the canonical \wt in the red and the solid line in Fig.
\ref{fig:w21.5} represents this power law, $\omega~=~2.3~\theta^{-0.8}$. After
transforming IP95 {\em F} band to m$_R$ of this work (F=m$_R$+0.094(B--R),
(B--R)=1.68, Metcalfe et al. 1991)  we find that the agreement for
$6''<\theta<68''$ is remarkable. However, at small angular scales
($\theta<6''$), except for the work by Carlberg et al. (1994), no previous work
shows a significant excess in \wt by more than one $\sigma$. 
The quality and amount of data that
we have used in this work are sufficient to make us believe that the
excess at small scales is a real effect.

\subsection{Comparison with Models}


How can a high clustering amplitude at small separations be understood? One of
the possibilities is to assume that mergers were more frequent at $z\approx0.35$
and the excess of pairs that we are seeing at that epoch is in fact mergers in
progress. They represent a true density excess. Another scenario that we cannot
rule out is the possibility of an excess of strongly clustered dwarf galaxies.
Two alternatives are possible in this hypothesis: a new class of dwarf
population proposed by Babul \& Rees (1992) which would be recently `active',
but subsequently fading out of view; or the `sub-units' proposed by Broadhurst
et al. (1992) which would slowly merge to form more massive galaxies.  It is
also possible that this excess in \wt is due to brightening of low luminosity
companions (Carlberg, 1992). A fourth possibility is that although there is
clustering evolution at scales larger than 6$''$ there is no clustering evolution
at scales $<6''$. In this case,  \wt is well fitted by a model with $\epsilon=0$
at $2''<\theta\le6''$ and $\epsilon=0.8$ at $\theta>6''$. Thus, there is   a
scale $\theta\approx6''$ ($\sim 20h^{-1}kpc$, $z\approx0.35$) at which the
clustering evolution slows down.

In order to show this latter point we have computed {\em standard
models} for \wt. If we assume a power law spatial correlation function,

\begin{equation}
\xi(r,z) = \left(\frac{r}{r_o}\right)^{-\gamma}(1+z)^{-(3+\epsilon)},
\end{equation}

\noindent 
where $r$ is the proper distance, $r_o$ is the proper correlation
length and $\epsilon$ is the clustering evolution index; $\epsilon=\gamma-3$
means clustering fixed in comoving coordinates and $\epsilon=0$ represents
stable clustering in physical coordinates. The angular correlation function is
obtained through Limber's equation (Peebles 1980 eqs [56.7] and [56.13] and
IP95). The no--evolution redshift distribution, $\frac{dN}{dz}$, is taken  from
Metcalfe et al. (1991) (see IP95 for discussion and justification of our
choice). 

In Fig. \ref{fig:w21.5} three models are shown: The {\em standard model}
($\gamma=1.8$, $\epsilon=0$, $\Omega=0.2$ and $r_o=5.1~h^{-1}Mpc$, solid line),
a model with clustering evolution, ($\gamma=1.8$, $\epsilon=0.8$, $\Omega=0.2$
and $r_o=5.1~h^{-1}Mpc$, dashed line), and a model with either  clustering
evolution, ($\gamma=1.8$, $\epsilon=-2.4$, $\Omega=0.2$ and $r_o=5.1~h^{-1}Mpc$,
dashed dotted line) or a model with a strong local correlation length
($\gamma=1.8$, $\epsilon=0$, $\Omega=0.2$ and $r_o=7.3~h^{-1}Mpc$, dash--dotted
line). 

The {\em standard model} with clustering  evolution ($\epsilon=0.8$)  fits very
well the observations in the range $6''\le\theta<24''$. The value of $\epsilon$
agrees with the results obtained by Shepherd et al. (1996), where the low
amplitude of the correlation function might be due to strong clustering
evolution since $z\approx0.35$. 

The small scale data ($\theta<6''$) is better fitted by a model with either a
high local clustering length ($r_o=7.3~h^{-1}Mpc$) or $\epsilon=-2.4$.   If we
assume a correlation length as observed for present day $L^*$ galaxies
($r_o=5.1~h^{-1}Mpc$), no clustering evolution ($\epsilon=0$), and no luminosity
evolution, the clustering properties of galaxies at $<z>=0.35$ and spatial
separations $\sim 20h^{-1}kpc$ (for $z\approx0.35$), in the $m_R$ passband are
similar to present day $L^*$ galaxies. The excess in \wt with respect to this
model might correspond to the galaxy merger rate discussed in the next section.

If the mean redshift of our sample is 0.35 then our data show a clear detection
of the scale, $\approx 19h^{-1}kpc$, at which the clustering evolution
approaches a highly non linear regime, where $\epsilon \le 0$ is expected.

\subsection{The galaxy merger rate}
 
The merger rate is expected to have a strong redshift dependence varying as
roughly as $(1+z)^m$, where $m \simeq 4.5 \Omega^{0.4}$. However, the difficulty
of identifying objects even at moderate redshift is a limiting factor in studies
of faint galaxy pairs. Another factor is imposed by the significant
amount of large telescope time required in order to obtain redshifts and good
quality images for a statistically significant sample of faint galaxies. So far,
only a few previous works have redshift information for faint galaxies in pairs
(see Koo \& Kron 1992 and Ellis 1995 for reviews on redshift surveys and imaging
of the faint galaxy population). 

There are two ways (not independent, though) to calculate the galaxy merger rate
from our data. First, we count the number of isolated pairs of galaxies in the
magnitude range $19\le m_R \le 21.5$, corresponding to $<z>\approx 0.35$, at scales
$2''\le\theta \le6''$ ($6.4h^{-1}kpc$ --  $19.3h^{-1}kpc$, for
$z\approx0.35$ and $q_o=0.1$),  and compare it with numbers of pairs from local galaxy surveys
selected in the same way. The fraction of galaxies in pairs in our catalog
is $10.04 \pm 0.17\%$ ($\sqrt{n}$ uncertainty). The
expected fraction of physical pairs is $0.636 \pm 0.007$,
if $\omega(2''-6'') = 1.75 \pm 0.05$.
Following Carlberg et al. (1994), the fraction of galaxies that will merge is
0.7. Thus, the percentage of galaxies in physical pairs that will eventually
merge is $10.04\times0.636\times0.7=4.47\pm 0.01\%$. At low redshifts
this value is $2.3\pm 0.2\%$ (Carlberg et al. 1994). If
the rate at which galaxies merge is proportional to $(1+z)^m$, then
$m=2.2\pm 0.5$. (The uncertainty in the merger rate is dominated by the
uncertainty in the fraction of galaxies that will merge. We assume 0.1.)

Second, the galaxy merger rate can be calculated using the excess correlation of
our faint galaxies over what would be expected from the {\em standard model}.
The fractional excess $\left({{\Delta N }\over{N}}\right)_{pairs}$ over what
would be expected from the {\em standard model} is $0.332$ ($<~\theta~>=
4''$ corresponding to $15.1h^{-1}kpc$ at $<~z~>=0.35$). Given that $(1+z)^m =
\frac{1+\omega_{observed}}{1+\omega_{standard}}$, then $m= 1.34$. It is
to be noticed that no corrections for the fraction of physical pairs that will
merge has been made in this case.

\section{Conclusions} 

We have estimated \wt from a well defined sample of galaxies at $19\le m_R \le
21.5$ and at small angular separation, $2''\le \theta \le 6''$. We have performed
simulations in order to test detection efficiency of galaxy pairs and  the
limiting magnitude in our galaxy sample. We find that our algorithms can detect,
at a 99\% confidence level, the two components of a pair separated by $> 2''$
and with $m_R < 22.5$ in our CTIO 4m prime focus CCD images. 

At $6''< \theta \le 68''$ the agreement between the present data and IP95 is
remarkable.  The amplitude of \wt at these separations is 1.6 times smaller than
the no-evolution standard models. 

Our analysis indicates that there is an excess of power in \wt at $2''\le \theta
\le 6''$ over what would be expected  from an extrapolation of the canonical
power law \wt at larger $\theta$. The significance of this excess  is $\approx
5\sigma$. These results place limits on the galaxy merger rate of
(1+z)$^{1.3-2.3}$.

At large angular scales ($6''< \theta \le 24''$) the data is best described by a
model where clustering evolution in $\xi(r,z)$ has taken place. Strong
luminosity evolution cannot be ruled out with the present data. At smaller
scales, $2''\le \theta \le 6''$, our data are formally fit by models where
$\epsilon=-2.4 (\Omega = 0.2)$ or the clustering length $r_o = 7.3h^{-1}Mpc$. If
the mean redshift of our sample is 0.35 then our data show a clear detection of
the scale ($\approx 19h^{-1}kpc$) where the clustering evolution approaches a
highly non linear regime, i.e., $\epsilon \le 0$.

\acknowledgments

We are grateful to the High-Z Supernovae Search Group for making their images
available. DFM thanks CNPq for the fellowship and LI thanks
Fondecyt Chile for support through `Proyecto 1960414'.

\clearpage

\begin{deluxetable}{cccccccccccc}
\tablecaption{Excess Number of pairs.
\label{tab:excess}}
\tablewidth{0pt}
\tablehead{ 
\colhead{Catalog } &  
\colhead{$\theta_1$--$\theta_2$} &   
\colhead{\wt}   & 
\colhead{$N_{total}$} & 
\colhead{$N_{pairs}^{obs}$} & 
\colhead{$N_{pairs}^{ran}$} & 
&&\colhead{Groups} &&&
\colhead{  $\left({{\Delta N }\over{N}}\right)_{pairs}$\tablenotemark{\dag}}\nl 
&
\colhead{[arcsec]}&
&&&&
\colhead{2} &  
\colhead{3} &  
\colhead{4} &  
\colhead{5} &  
\colhead{6} &
} 
\startdata

 
Real&      2 --   6& 1.753& 16749& 1317& 488& 841& 114& 15& 4& 3& 0.361 \nl
Random&    2 --   6&-0.010& 16749&  477& 497& 438&  14&  0& 0& 0&  --    \nl
\nl
Real&      6 --  24& 0.250& 16749& 9962& 7921& 1318& 314& 58& 13& 2&-0.011\nl
Random&    6 --  24& 0.004& 16749& 7970& 7943& 1587& 255& 30& 5& 1& -- \nl
 
\enddata
\tablenotetext{\dag}{Fractional excess number of pairs over what would be expected from
IP95.}
\end{deluxetable}

\clearpage

\figcaption [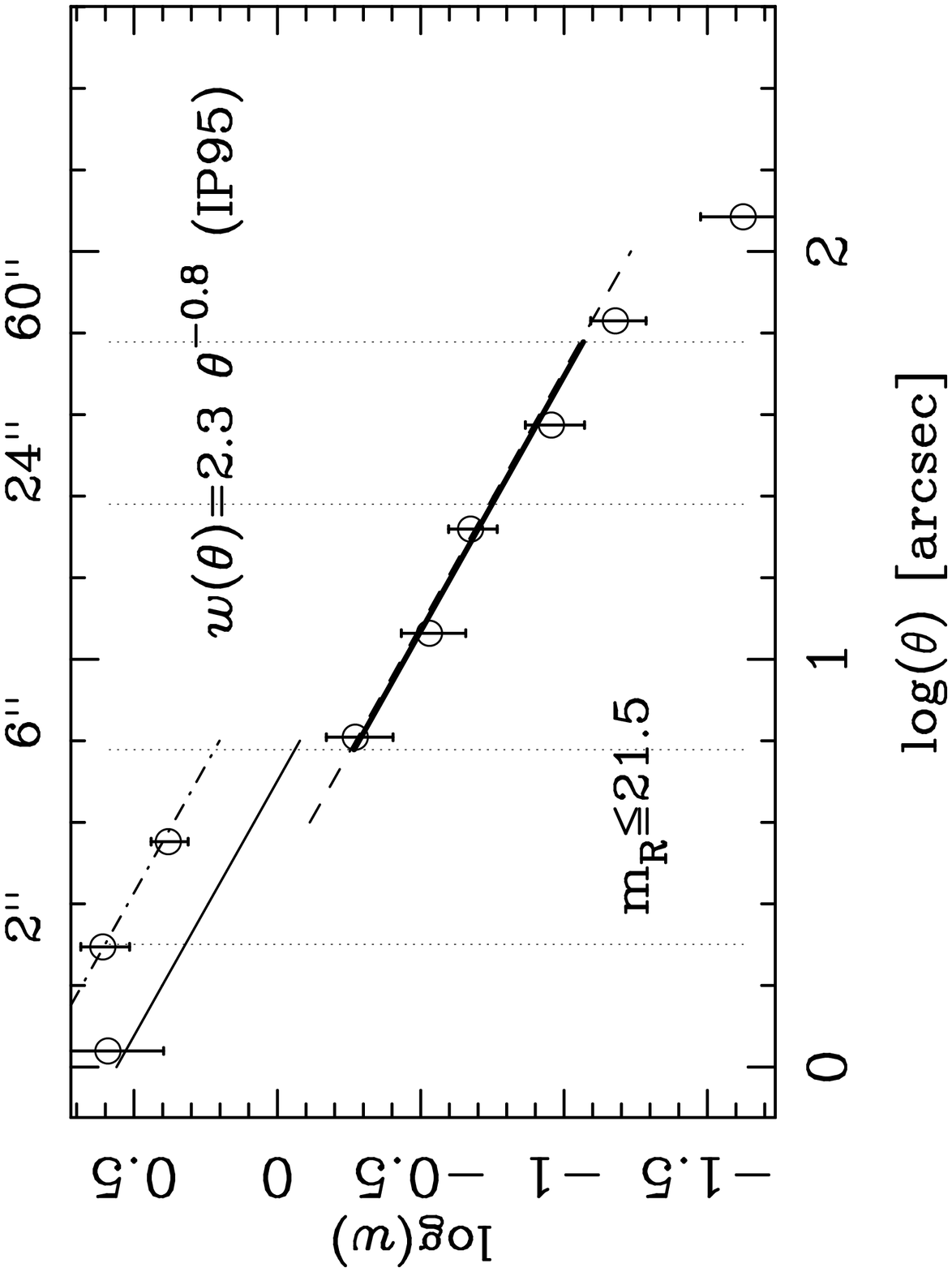]{ Angular Correlation Function for $19\le m_R \le 21.5$.
The open circles are the data. 65\% uncertainties (500 bootstrap resampling
of the data) are shown as vertical lines. 
The dark solid line is the IP95 \wt result with $\delta=-0.8$.
The thin solid line is the no-evolution model described in the text 
($r_o=5.1~h^{-1}$ Mpc, $\epsilon=0$). 
The dashed line is a  model with clustering evolution 
($r_o=5.1~h^{-1}$ Mpc, $\epsilon=0.8$).
The dashed dotted line is a  model with either clustering evolution 
($r_o=5.1~h^{-1}$ Mpc, $\epsilon=-2.4$) or strong correlation length ($r_o=7.3~h^{-1}$ Mpc,
$\epsilon=0$).
\label{fig:w21.5}} 

\figcaption [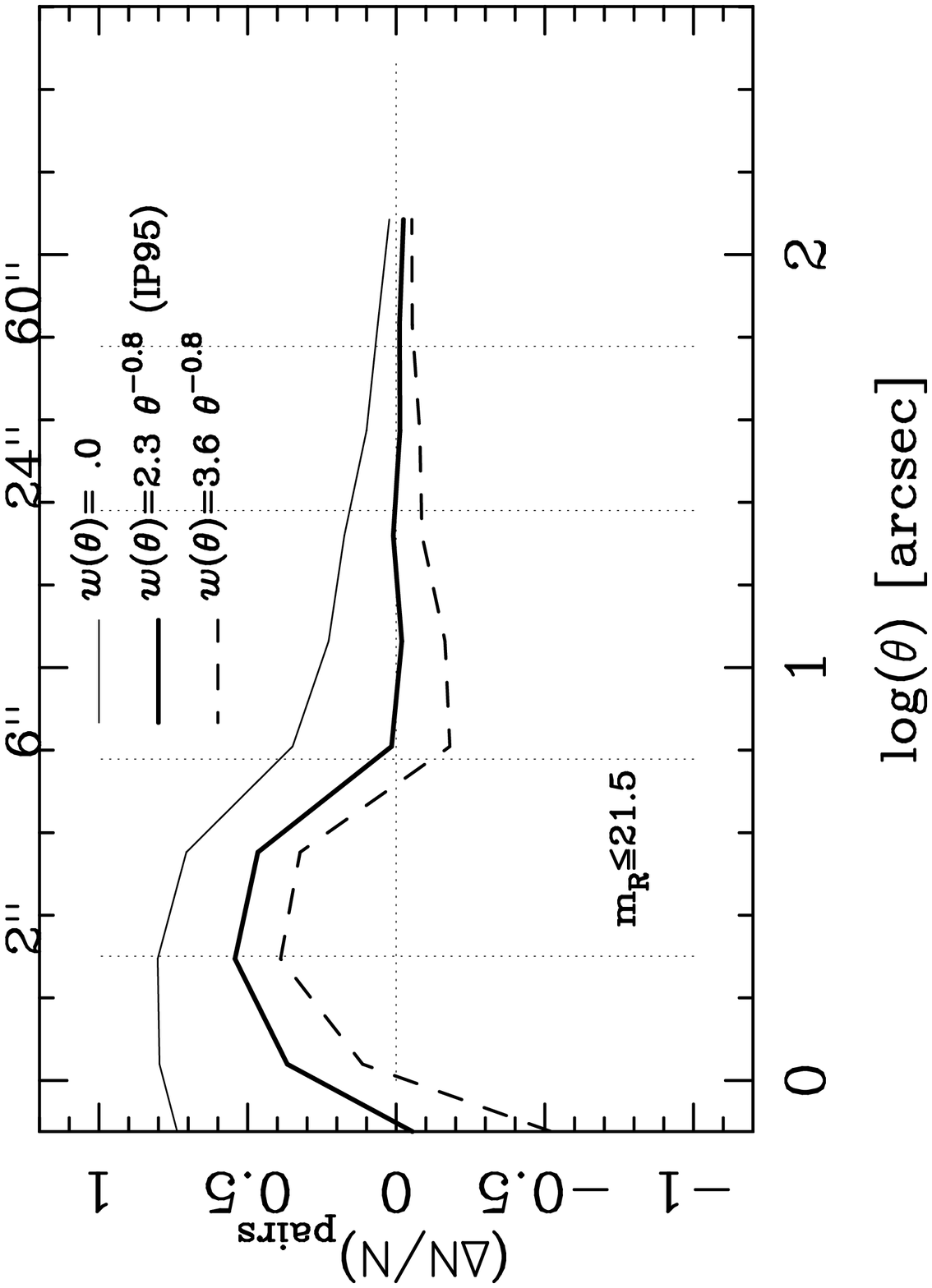]{ Fractional excess number of pairs for $19\le m_R \le
21.5$, as defined in equation 2. The dark solid line is the IP95 \wt result with $\delta=-0.8$.
The dashed line is the no-evolution model described in the text 
($r_o=5.1~h^{-1}$ Mpc, $\epsilon=0$). 
 The thin solid line is the fractional excess with respect to a random
distribution of points.
\label{fig:dn21.5}}

\clearpage

\plotone{w21.5.ps}

\plotone{dn21.5.ps}

\end{document}